\documentclass{article}
\usepackage{spconf,amsmath,graphicx,hyperref,amssymb,array,bookmark,tabularx,float}
\usepackage{multirow}
\usepackage{booktabs}
\usepackage{siunitx}
\usepackage{caption}
\title{EMOANTI: AUDIO ANTI-DEEPFAKE WITH REFINED EMOTION-GUIDED REPRESENTATIONS}
 \name{Xiaokang Li \qquad Yicheng Gong$^{\dagger}$  \qquad Dinghao Zou \qquad Xin Cao \qquad Sunbowen Lee
 \thanks{
  \begin{tabular}{@{}l@{}}
  $^\dagger$Corresponding author \\
  \textsuperscript{1}https://anonymous.4open.science/r/EmoAnti
  \end{tabular}
  }}
   \address{Wuhan University of Science and Technology\\
   College of Science\\
   Wuhan,China}
\begin{document}
%
\maketitle
\begin{abstract}
Audio deepfake is so sophisticated that the lack of effective detection on methods is fatal.
While most detection systems primarily rely on low-level acoustic features or pretrained speech representations, 
they frequently neglect high-level emotional cues, which can offer complementary and potentially anti-deepfake information to enhance generalization. 
In this work, we propose a novel \textbf{audio anti-deepfake} system that utilize \textbf{emotional features(EmoAnti)} by exploiting 
a pretrained Wav2Vec2 (W2V2) model fine-tuned on emotion 
recognition tasks,which derive emotion-guided representations, 
then designing a dedicated feature extractor based 
on convolutional layers with residual connections
to effectively capture and refine emotional characteristics 
from the transformer layers outputs.
Experimental results show that our proposed architecture achieves state-of-the-art performance on both the ASVspoof2019LA and ASVspoof2021LA benchmarks, and demonstrating strong generalization on the ASVspoof2021DF dataset.
Our proposed approach's code is available at Anonymous GitHub\textsuperscript{1}.
\end{abstract}
\section{Introduction}
\label{sec:intro}
Text-to-speech (TTS) and voice conversion (VC) technologies, empowered by the rapid advancement of deep learning, have achieved remarkable realism in synthesized speech generation~\cite{yi2023audio}.
 While this progress has expanded their application domains, it also raises significant security concerns: such synthetic speech can be exploited to launch spoofing attacks against automatic speaker verification (ASV) systems, 
 potentially leading to financial loss and privacy breaches. This has driven increasing research interest in audio anti-deepfake as a critical countermeasure.

 Early audio anti-deepfake methods primarily relied on low-level acoustic features such as fundamental frequency (f0)~\cite{xue2022audio} and LFCC~\cite{arif2021voice}. 
 For example, Xue et al.~\cite{xue2022audio} combined f0 with real and imaginary spectrograms for anti-deepfake. While interpretable, these approaches often neglect high-level emotional information and exhibit limited generalization.
 Leveraging emotional features, Conti et al.~\cite{conti2022deepfake} proposed a neural network to detect audio deepfake, whose current generation techniques are unable to accurately synthesize natural emotional behaviors in speech. To extract key features from deepfake audio, many approaches~\cite{martin2022vicomtech,zhang2023audio,guo2024audio} use self-supervised speech models, which learn general-purpose representations from  large-scale unlabeled speech, as their backbone. 
 Although these approaches achieve promising performance, they lack the ability to interpret the general-purpose features learned by the pretrained models.

To address both the interpretability of learned representations and the generalization capability of the model, we propose a novel audio anti-deepfake framework: EmoAnti. 
\textbf{Our approach is built upon two key components: 1) to capture high-level emotional information while enhancing the interpretability of speech embeddings, we adopt Wav2Vec2 as the backbone model and fine-tune it on a speech emotion recognition (SER) task; 
2) to improve generalization, we design a convolutional residual feature extractor(CRFE) to refine emotion-guided representations from the transformer layers of Wav2Vec2 and identify subtle emotional discrepancies between bonafide and spoof speech.}

 Our contributions are as follows:

 1) Based on emotional-aware embeddings, we propose a anti-deepfake method, which achieves strong generalization performance and maintains interpretability.

 2) To better capture and refine emotion-guided representations in speech, we design a convolutional residual feature extractor built upon Wav2Vec2.

 3) To validate the importance of emotional information in audio anti-deepfake, we conduct comprehensive ablation studies, which provide new insights for future research.
\begin{figure*}[t]
  \centering
  \includegraphics[width=\linewidth]{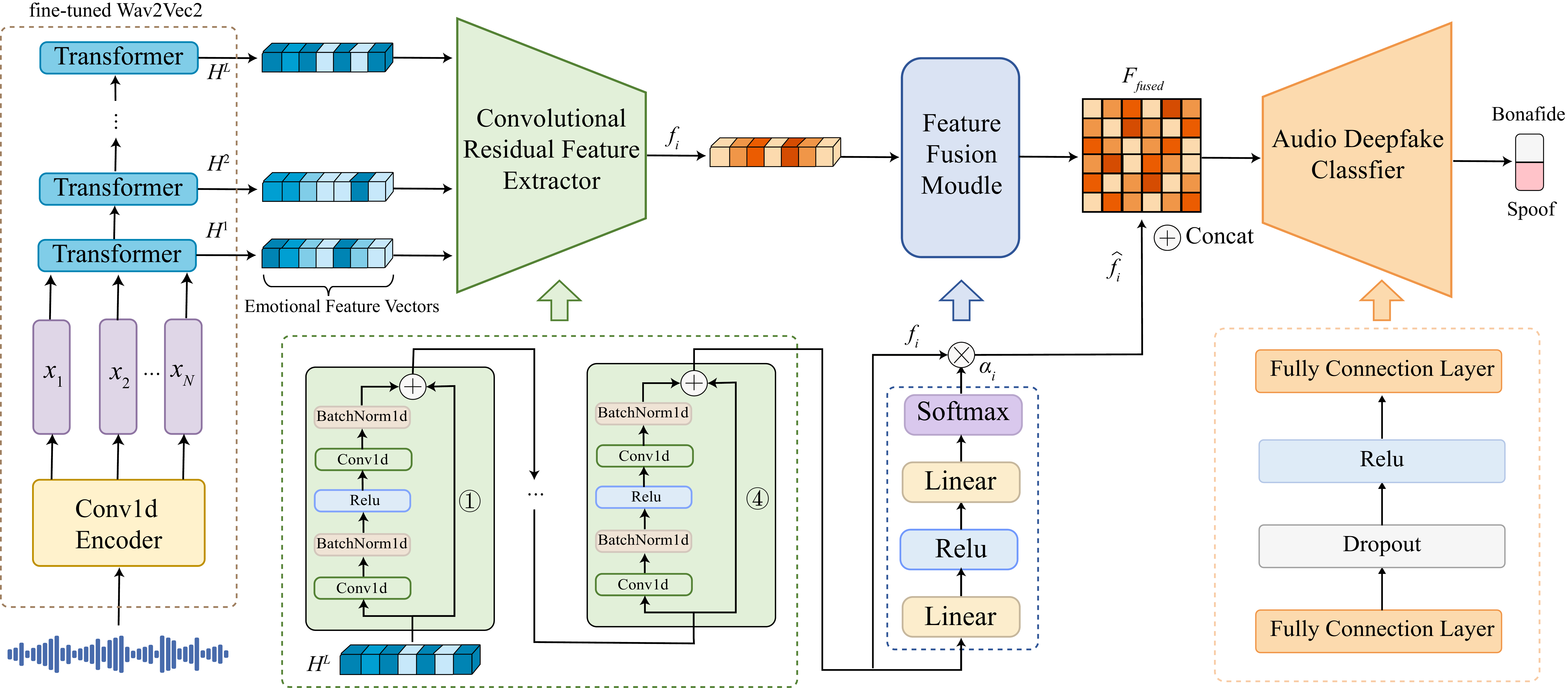} 
  \caption{Overview of the EmoAnti framework.Dashed - line parts denote the internal detailed structures and the workflow during model operation.}
  \label{fig:architecture}
\end{figure*}
  
\section{PROPOSED METHOD}
\label{sec:proposed_method} 
The overall architecture of the proposed model is illustrated in Fig.~\ref{fig:architecture}. Given raw audio waveforms as input, the fine-tuned Wav2Vec2 model processes them through a series of Conv1d and transformer layers to extract high-level emotion-guided representations. These features are then passed to the convolutional residual feature extractor and feature fusion moudle, which further refines and fuses multi-level emotional cues to generate enhanced fusion feature vectors. Finally, the fused representations are fed into a classifier to generate the final prediction. This section provides a detailed description of each component in the framework.

\subsection{Fine-tuned Wav2Vec2 Model}
In this study, we adopt the wav2vec2-large-robust model \cite{hsu2021robust,wav2vec2_large_robust} as the frontend encoder. It consists of three main components: 1) a 7-layer convolutional feature extractor that transforms raw waveforms into a feature sequence \(X = \{x_1, x_2, \cdots, x_N\}\) using strided convolutions—initially with a stride of 5, followed by strides of 2—thereby reducing the temporal resolution; 2) a 12-layer transformer encoder (with 16 attention heads per layer) that captures long-range temporal dependencies; 3) a quantization module for latent speech representation learning. Let  \(H^l = \{h^l_1, h^l_2, \cdots, h^l_N\}\) denote the output of the \textit{l} -th transformer layer (\(l \in [1, 12]\)), and let  \(H = \{H^l\}_{l=1}^{12}\) represent the collection of outputs from all 12 transformer layers.

\begin{figure}[ht]
  \centering
  \includegraphics[width=0.8\linewidth]{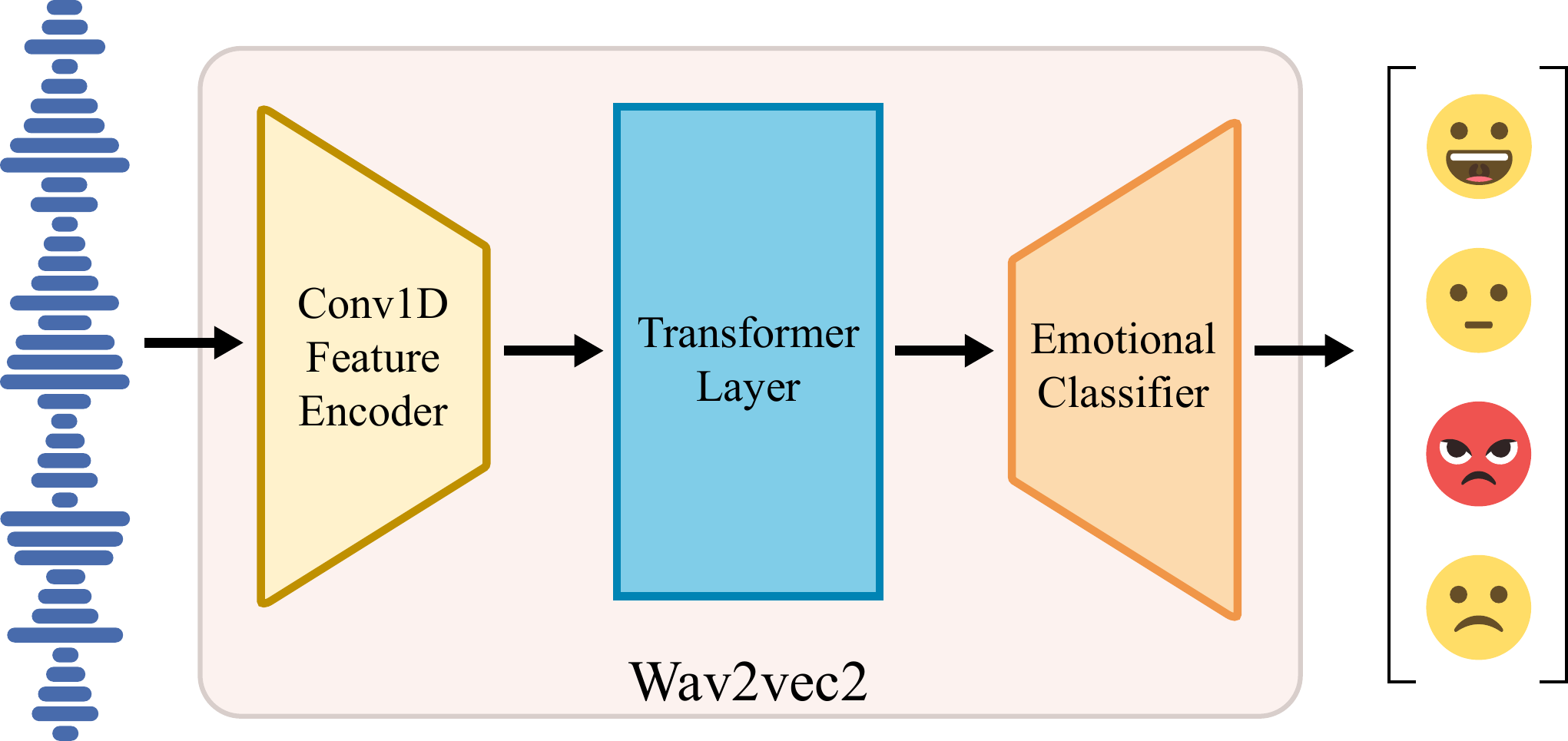} 
  \caption{Fine-tuning for four-category emotion recognition (happy, neutral, angry and sad from top to bottom, corresponding to the labels).}
  \label{fig:fine-tuning}
\end{figure}

To adapt the pretrained model for emotion recognition and audio anti-deepfake, we fine-tune it by retaining the backbone of wav2vec2-large-robust—including the convolutional feature extractor and transformer encoder—to leverage its latent speech representations, while adding task-specific classification heads. For emotion recognition (Fig.~\ref{fig:fine-tuning}), the classification head takes the final transformer layer’s output $H^L$ as input and predicts emotion categories by leveraging paralinguistic cues. 
This enables the model to effectively transfer the emotion-guided representations learned during pretraining to the task of audio anti-deepfake.
\subsection{Convolutional Residual Feature Extractor}
While final-layer representations of pre-trained transformers suffice for speech recognition, prior work shows that earlier and intermediate layers yield more discriminative features for speaker and paralinguistic tasks~\cite{chen2022large, wagner2023dawn}. Inspired by this, we design a convolutional residual feature extractor to aggregate and refine emotional representations from multiple transformer layers, thereby enhancing the ability to distinguish subtle emotional details in audio deepfakes.

Specifically, the module consists of four sequential convolutional residual blocks.
Each block's main branch contains two Conv1d layers (kernel size = 3, padding = 1): 
the first projects the input dimension $d_{\text{in}}$ to a hidden dimension $d_{\text{hidden}}$ through BatchNorm1d and ReLU activation, while the second preserves $d_{\text{hidden}}$ with an additional BatchNorm1d layer.

\begin{equation}
  \begin{aligned}
  h_{\text{conv1}} &= \mathrm{BNorm1d}\bigl(\mathrm{Conv1d}_{3,\text{pad}=1}(H^l)\bigr), \\
  h_{\text{conv2}} &= \mathrm{BNorm1d}\Bigl(\mathrm{Conv1d}_{3,\text{pad}=1}\bigl(\mathrm{ReLU}(h_{\text{conv1}})\bigr)\Bigr),
  \end{aligned}
  \label{eq:conv_branch}
  \end{equation}
  where \(H^l\) is the input hidden representation from the \(l\)-th transformer layer.

The residual branch uses a \(1 \times 1\) Conv1d if \(d_{\text{in}} \neq d_{\text{hidden}}\), otherwise it applies identity mapping:

\begin{equation}
  h_{\text{residual}} =
  \begin{cases}
  \mathrm{Conv1d}_{1}(H^l), & d_{\text{in}} \neq d_{\text{hidden}},\\
  H^l, & d_{\text{in}} = d_{\text{hidden}}.
  \end{cases}
  \label{eq:residual_branch}
  \end{equation}
The block output is then obtained by element-wise addition and ReLU activation:
\begin{equation}
  f = \mathrm{ReLU}\bigl(h_{\text{conv2}} + h_{\text{residual}}\bigr),
  \label{eq:fusion_output}
  \end{equation}
followed by transposing \(f\) to \([B, T, d_{\text{hidden}}]\) for storage and further use.

\subsection{Feature Fusion Module}   
To effectively aggregate the refined emotion-guided representations from the CRFE, we introduce a feature fusion module that applies a dedicated temporal attention subnetwork to each output feature \( f_i \) generated by the convolutional residual blocks, enabling adaptive weighting of temporal segments across different levels of abstraction.
\begin{equation}
  \begin{aligned}
  e_{i,t} &= \text{Linear}_2\bigl(\mathrm{ReLU}(\text{Linear}_1(f_{i,t}))\bigr), \\
  \alpha_{i,t} &= \frac{\exp(e_{i,t})}{\sum_{t'=1}^{T}\exp(e_{i,t'})}, \\
  \hat{f}_i &= \sum_{t=1}^T \alpha_{i,t} \odot f_{i,t},
  \end{aligned}
  \label{eq:attention_aggregate}
  \end{equation}

where \( f_{i,t} \) denotes the \(t\)-th temporal segment of \(f_i\) and \( e_{i,t} \) denotes the attention score, \( \alpha_{i,t} \) is its normalized weight, and \( \hat{f}_i \in \mathbb{R}^{d_{\text{hidden}}} \) is the temporally aggregated feature. 

Finally, we concatenate all \( \hat{f}_j \) (\(j \in [1, i]\))into the fused feature:

\begin{equation}
F_{\mathrm{fused}} = [\hat{f}_1,\, \hat{f}_2,\, \dots,\, \hat{f}_i].
\label{eq:fused_feature}
\end{equation}

\subsection{Audio Deepfake Classifier} 
This module takes the fused feature matrix $\mathbf{F}_{\text{fused}} \in \mathbb{R}^{B \times D_{\text{total}}}$ as input, where $B$ denotes the batch size and $D_{\text{total}}$ represents the total dimensionality of the fused features. It performs classification through a series of linear transformations, non-linear activations, and regularization mechanisms. Mathematically, the entire process can be expressed as:
\begin{equation}
  y_{\text{logits}} = \text{Linear}_2\Big(\text{ReLU}(\text{Dropout}(\text{Linear}_1(F_{\text{fused}})))\Big)
\end{equation}
  
\section{EXPERIMENTS}
\label{sec:experiments}
\subsection{Datasets and Evaluation Metrics} 
We fine-tune the model on IEMOCAP~\cite{busso2008iemocap}, following the emotion label selection in prior work~\cite{morais2022speech}: "angry", "happy" (including "excited"), "sad", and "neutral", resulting in 5,531 utterances. Performance is evaluated using accuracy and macro F1-score.

For anti-deepfake, we evaluate on ASVspoof2019 LA and ASVspoof2021 LA/DF~\cite{todisco2019asvspoof,yamagishi2021asvspoof,liu2023asvspoof}. Training is performed on the ASVspoof2019, as ASVspoof2021 provides no new training data. The ASVspoof2021 are more challenging, with 2021 LA introducing new speakers and transmission artifacts, while 2021 DF targets compressed deepfake attacks. Evaluation metrics are the minimum normalized tandem detection cost function (min t-DCF)~\cite{kinnunen2018t} and equal error rate (EER).

\subsection{Training Setup} 
During fine-tuning, the Wav2Vec2 model was optimized using the AdamW optimizer (learning rate = $1 \times 10^{-5}$, weight decay = 0.01), with a cosine annealing learning rate schedule and 10\% warm-up. Gradient accumulation was employed to simulate larger effective batch sizes. Model selection was based on the macro F1 score on a validation set, with checkpoints evaluated periodically and the best models retained for downstream tasks. Training was performed for 8 epochs.

For audio anti-deepfake, we used the Adam optimizer (learning rate = $1 \times 10^{-4}$) and CrossEntropyLoss. Models were trained for 6 epochs with evaluation after each epoch, and the checkpoint achieving the lowest validation loss was preserved. All experiments were conducted with dataset-specific random seeds: 43, 44, 45 for ASVspoof2019 LA; 43, 45, 456 for ASVspoof2021 LA; and 46, 47, 78 for ASVspoof2021 DF.
\begin{table}[ht]
  \centering
  \caption{Comparison results of EER (\%) and min t-DCF between our proposed method and other anti-deepfake systems on the ASVspoof2019LA evaluation set.}
  \label{tab:asvspoof_2019}
  \resizebox{\linewidth}{!}{ 
    \begin{tabular}{
      l  
      S[table-format=2.2]  
      S[table-format=1.4]  
    }
    \toprule
    \textbf{System} & {EER (\%)} & {min t-DCF} \\
    \midrule
    LFCC-GMM\cite{asvspoof2021_baseline_cm_2021} & 8.09  & 0.212  \\
    CQCC-GMM\cite{asvspoof2021_baseline_cm_2021} & 9.57  & 0.237  \\
    LFCC-LCNN\cite{das2021known} & 5.06 & 0.237  \\
    RawNet2\cite{tak2021end} & 5.64    & 0.130  \\
    ECAPA-TDNN\cite{chen2021ur} & 4.58 & 0.117 \\
    W2V2 (fixed)+LCNN+BLSTM\cite{wang2021investigating} & 1.47 & 0.105 \\
    W2V2 (finetuned)+LCNN+BLSTM\cite{wang2021investigating} & 2.31 & 0.120 \\
    \midrule
    \textbf{EmoAnti (ours)} & \textbf{0.44} & \textbf{0.0139} \\
    \bottomrule
    \end{tabular}
  }
\end{table}

\section{Results and Analysis}
\label{sec:results}
\subsection{Results on ASVspoof2019 LA}
Table~\ref{tab:asvspoof_2019} compares our method with representative approaches that use either handcrafted low-level acoustic features or self-supervised speech models as frontends. On the ASVspoof2019LA evaluation set, our model achieves state-of-the-art performance with an EER of 0.44\% and a min t-DCF of 0.0139.

Among these methods, those based on handcrafted low-level features (e.g., LFCC, CQCC) exhibit limited performance, while approaches leveraging pretrained models such as Wav2Vec2 show significant improvements. 

Our approach further enhances representation learning by fine-tuning the pretrained model on emotion recognition and refining the resulting representations through the CRFE, enabling it to capture subtle high-level emotional variations in speech. This strong performance demonstrates the effectiveness of modeling emotional discrepancies for anti-deepfake. Our results suggest that leveraging refined emotional cues not only improves detection accuracy but also opens a promising new direction for research in audio anti-deepfake.

\begin{table}[ht]
  \centering
  \caption{Results of EER (\%) and min t-DCF comparison between our proposed method and other anti-deepfake systems on the ASVspoof2021LA and DF evaluation sets.}
  \label{tab:asvspoof_2021}
  \resizebox{\linewidth}{!}{ 
    \begin{tabular}{
      l
      S[table-format=2.2]
      S[table-format=1.4]
      S[table-format=2.2]
    }
      \toprule
      \multirow{2}{*}{\textbf{System}} & \multicolumn{2}{c}{\textbf{LA}} & \textbf{DF} \\
      \cmidrule(lr){2-3}
      & {EER (\%)} & {min-t-DCF} & {EER (\%)} \\
      \midrule
      LFCC-GMM\cite{asvspoof2021_baseline_cm_2021} & 19.30 & 0.5760 & 25.25 \\
      CQCC-GMM\cite{asvspoof2021_baseline_cm_2021} & 15.62 & 0.4970 & 25.56 \\
      LFCC-LCNN\cite{das2021known} & 9.26 & 0.3450 & 23.48 \\
      RawNet2\cite{tak2021end} & 9.50 & 0.4260 & 22.38 \\
      ECAPA-TDNN\cite{chen2021ur} & 5.46 & 0.3094 & 20.33 \\
      W2V2 (fixed)+LCNN+BLSTM\cite{wang2021investigating} & 10.97 & 0.4720 & 7.14 \\
      W2V2 (finetuned)+LCNN+BLSTM\cite{wang2021investigating} & 7.62 & 0.3770 &  \textbf{5.44} \\
      \midrule
      \textbf{EmoAnti (ours)} & \textbf{4.62} & \textbf{0.2920} & 13.72 \\
      \bottomrule
    \end{tabular}
  }
\end{table}

\subsection{Results on ASVspoof2021 LA and DF}
Table~\ref{tab:asvspoof_2021} presents the performance of our method and baseline approaches on the LA and DF of the ASVspoof2021 dataset. 
On the 2021LA, our model achieves the best results with an EER of 4.62\% and a min t-DCF of 0.292. On the 2021DF, it achieves an EER of 13.72\%, outperforming most models. 
These results demonstrate the effectiveness and robustness of the proposed method across challenging evaluation conditions.

\begin{figure}[H]  
  \centering  
  \includegraphics[width=\linewidth]{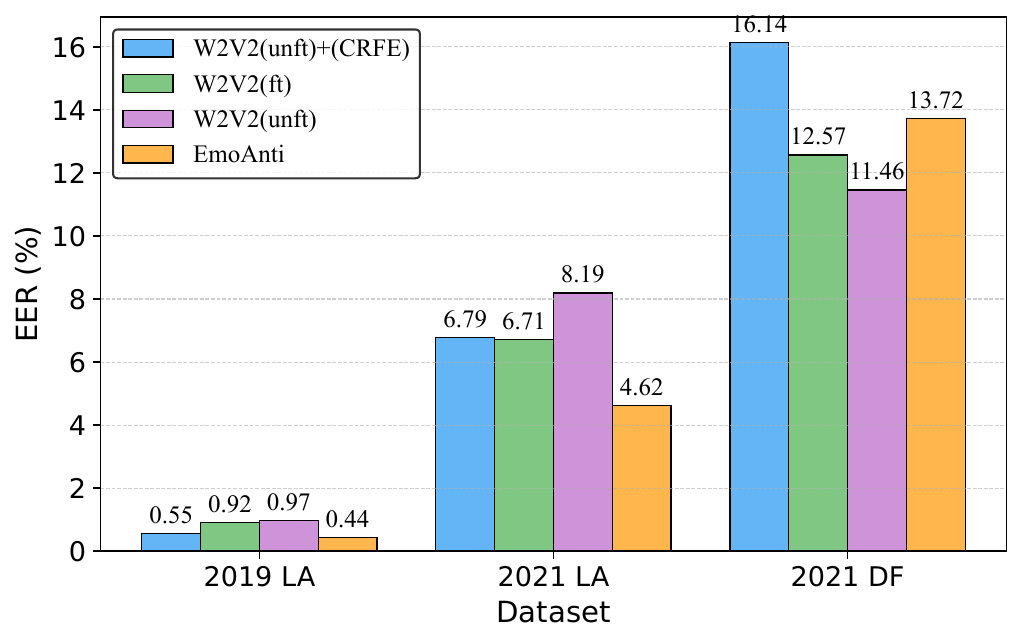}  
  \caption{Ablation Study Results (EER\%). \textit{ft}:fine-tuned, \textit{unft}:unfine-tuned, \textit{CRFE}:convolutional residual feature extractor.}  
  \label{fig:ablation}  
\end{figure}
\subsection{Ablation Study}
Fig.~\ref{fig:ablation} summarizes our ablation studies with three experiments: 1) removing emotion-aware fine-tuning; 2) removing CRFE while keeping fine-tuning; 3) removing both. 

On the ASVspoof2019 LA dataset, EmoAnti performing the best at 0.44\%. On ASVspoof2021 LA, EmoAnti again outperforms other models, achieving an EER of 4.62\%. Notably, on the 2021 DF dataset, removing only the emotion-aware fine-tuning yields the worst performance with an EER of 16.14\%, whereas removing both fine-tuning and the CRFE results in the best performance at 11.46\% EER.

These results demonstrate that the coupling of emotion-aware fine-tuning and the CRFE is highly effective for audio deepfake detection. The observed performance trend on the 2021 DF dataset may due to the model overfit to characteristics of the Logical Access (LA) data or the CRFE tends to capture fine-grained local features at the expense of global contextual information, potentially compromising generalization to compressed and transcoded speech.

\section{CONCLUSION}
In this paper, we propose EmoAnti for audio anti-deepfake, which is evaluated on three datasets. The main contribution of EmoAnti is constructing a Wav2Vec2 fine-tuned on emotion recognition tasks to extract emotional cues from speech, with subsequent refinement via a convolutional residual feature extractor to distinguish emotional differences between spoofed and bonafide speech.
Comparative experiments demonstrate that our approach leveraging emotional details achieves excellent performance in audio anti-deepfake. Meanwhile, ablation studies further validate the significance of emotional information for anti-deepfake. There is still room for improvement: we can replace the CRFE with a model that attends to global information to enhance generalization on the 2021DF dataset.

\section{ACKNOWLEDGMENTS}
This work was supported by the National Natural Science Foundation of China (NO.12171378) and High Performance Computing Center of Wuhan University of Science and Technology. We thank the reviewers for their contributions.

\AtBeginEnvironment{thebibliography}{
  \baselineskip=10pt 
}
\bibliographystyle{IEEEbib}
\bibliography{EmoAnti}

\end{document}